\documentclass[%
 letterpaper,
 reprint,
 showpacs,
 showkeys,
nofootinbib,
 amsmath,amssymb,amsfonts,
 aps,
 pra,
]{revtex4-1}

\usepackage{graphicx}
\usepackage{dcolumn}
\usepackage{amsmath}
\usepackage{array}
\usepackage{bm}
\usepackage{hyperref}

\begin{document}

\title{``Time operator'': the challenge persists}

\author{Bogdan Mielnik}
\email{bogdan@fis.cinvestav.mx}

\author{Gabino Torres-Vega}
\email{gabino@fis.cinvestav.mx}

\affiliation{Departamento de F\'\i sica, Cinvestav, AP 14-740, 07000,
  M\'exico D.F., M\'exico.}

\begin{abstract}
Contrary to the conviction expressed by J.~Kijowski [Phys. Rev. A
{\bf 59}, 897 (1999)] and shared in some other papers, the reasons
to look for the `time operator' in the context of the standard quantum
doctrine of orthogonal projectors and self-adjoint observables are
highly questionable. Some improved solutions in terms of POV
measures invite critical discussions as well. This paper appeared as {\em
Concepts of Physics\/ \bf 2}, 81 (2005).
\end{abstract}


\maketitle

The time of a quantum event is a random variable. The fact inspired a
patient quest for the {\it time operators}
\cite{Aha61,Ki74,Grot,De97,Gian97,Muga98,De99,Unru99,Koch99,Leav00},
originally following the orthodox idea of quantum observables
\cite{Dir,Neu}, subsequently reformulated in terms of positive operator
valued (POV) measures. In some ocassions, when an operator was born with an
anatomic defect, it was corrected by a `self--adjoint surgery' and a {\it final
solution\/} was announced (see also cf. \cite{Unru99}). In his essay of 1999, J.
Kijowski \cite{Ki99} closes the cycle by announcing that the final solution was
known indeed since 1974 \cite{Ki74}. Strangely, his comment seems like a
step back toward the most traditional orthodoxy, comparing even with his 1974
paper. The {\em observables\/} are again represented by self-adjoint
operators, the {\em probabilities\/} given by the spectral measures, etc. The
parts of Kijowski's solution had been adopted by other authors, though were
differently composed. In this comment we try to find out whether the Kijowski
solution, or some reformulated proposals can be indeed conclusive.

We start from the orthodox trend. The first attempts at constructing  
the time operator in the traditional frame of Dirac and v. Neumann
\cite{Dir,Neu} (circumventing the Pauli theorem \cite{Pauli}), pretended
to check the universality of the existing formalism. The next attemts, though
non-relativistic, might have an additional sense: to keep the space and time 
variables on equal footing, preparing the ground for the hypothetical space-time
quantization \cite{Grot,rove,reis}.

The resulting {\em time operators\/} show some basic similarities. For the free
particle, they all depart from the classical formula $t = -q/p$ (where $q$ and
$p$ are the position and momentum, we put $m=1$); the differences consist in
methods ``to make them self-adjoint''. In case of Kijowski \cite{Ki74,Ki99},
the operator is constructed in two steps: (1) the kinetic energy operator
$H=p^2/2$ of Schr\"odinger's quantum mechanics in 1 space dimension is replaced
by the {\it pseudoenergy}
\begin{equation}
\Xi=\Xi(p) = \rm{sgn}(p){p^2\over 2} ={1\over2} p|p|\quad,
\label{xi}
\end{equation}
with continuous eigenvalues $\xi\in\mathbb{R}$; and (2) the {\it pseudotime\/} 
$\Theta$ (with continuous eigenvalues $\theta\in\mathbb{R}$) is defined as the
operator
canonically conjugate to $\Xi$ 
\begin{equation}
\Theta = - i {\partial\over\partial\xi}\;.
\label{theta}
\end{equation}

Following the orthodox rules, Kijowski applies the standard axioms originally
formulated for the instantaneous measurements \cite{Dir}. Thus, he uses the
spectral projectors $P(\Delta)$ of $\Theta$ on the intervals $\Delta\subset
\mathbb{R}$ in a traditional way, to define the probabilities
$p_\phi(\Delta)=\langle\phi|P(\Delta)|\phi\rangle$ of the particle arrival in
the time intervals $\Delta$ for any normalized initial state $\phi$ (cf.
\cite{Ki99}, p. 898, formula (5)). Note, that the original approach of
\cite{Ki74} was modified in \cite{Grot,De97}, then modernized in
\cite{Gian97,Muga98,De99} and subsequent papers (cf.
\cite{Leav00,Muga01,Muga02})
in terms of the POV probability measures. Due to its nice self-adjoint form,
the Kijowski operator $\Theta$ (and its descendants) are close to winning the
competition for the best ``time of arrival''. However, before comparing the
merits of various formal constructions solving the problem, we would like to
know {\em what is precisely the problem which they solve?}

Like all other observables, the {\em time operator\/} should represent some
measurement. Several types of physical experiments might be  intuitively 
associated with the concept of ``time'', and if these intuitions 
differ, so may the operators. Unfortunately, a realistic background is almost
absent in most papers on the time operator, written in almost Aristotelean
spirit. 

A widely shared idea seems to be that the {\em time measurement\/} is 
performed by a {\em waiting  detector,\/}  programmed to register  a  certain 
definite event (e.g., the arrival of a microobject to the sensitive part 
of the apparatus). One of simplest such devices is a flat, macroscopic 
screen, producing a sharp, brilliant spark when hit by a microobject. Is 
this an adequate physical model behind the ``time  operator''? Not necessarily;
some questions are still open and they can make a lot of difference! 

In the first place, should the screen register the particles arriving from one
or from both sides? A natural scenario would be the one-sided screen
\cite{BM94,March} (a microparticle approaches to the screen from its sensitive
side, to be registered at some moment $t$). This, of course, involves the
assumption that the screen is impenetrable. However, the idea does not seem to
prevail (except perhaps Marchewka and Schuss \cite{March}). Anyhow, it is not
stated by Kijowski; nor by Grot et al\cite{Grot}; neither we find it in majority
of papers on the time of arrival. Let us therefore adopt an opposite view: we
shall consider a {\em two sided\/} screen, which does not discriminate
half-space, and is sensitive to particles crossing in both directions.  

If so, the next question is: can the particle sneak across the screen 
undetected? Presumably so (should the screen preclude the tunneling, we would
return to the previous option). In fact, the image of particle circulating on
both sides of the screen seems implicit in almost all constructions of the
``time operators'' and POV measures. Enough to look at the proposed time
probability distributions (see an interesting review \cite{Leav00}); the authors
don't care to split the packet into two space separated, non interfering
components on two sides of the screen, but they care a lot to split it into the 
positive and negative momentum parts ({\em right\/} and {\em left  
movers}). This is of course not the same.  

The next question is less trivial: is the evolution of the particle 
before the moment of ``arrival'' affected by the existence of the screen? 
The problem seems crucial in the approach of Kijowski and other authors trying
to construct the time measurement either in terms of projectors or POV-measures.
 When reading \cite{Ki74} one might guess that the measuring apparatus is a
maximally {\it non-intrusive\/} device, which waits inconspicuously,
without perturbing the particle until the moment when the particle is detected.
So, is it completely transparent? Does the particle obey the free evolution
until detected? Such a hypothesis seems to emerge from Kijowski comment (see
\cite{Ki99}, p. 898 left, l.7-5 from bottom), though his formulation is a bit
enigmatic: ``On the other hand'', he writes, ``any quantum state (\dots)
undergoes the {\it standard\/} `chronological evolution' from time $t_1$ to
$t_2$, described by the Schr\"odinger equation''. However, what does it mean?
Does the particle follow an equation which takes into account the influence of
the screen? Or does it obey the {\it free Schr\"odinger equation\/} until
detected? The last option is visibly privileged in \cite{Ki74}, Ch. 5, p. 367,
where the probability density on the time axis is constructed out of the {\it
freely evolving\/} wave packet $\phi_t$. The fact is considered so obvious that
the formula $A(t)= F(\phi_t)$ on p. 367 is not even included into the list of
formally stated axioms. The same idea, apparently, is shared by other authors
who develop the POV formalism\cite{Muga98,De99,Leav00}. Yet, obvious it is not! 

If the particle survived without being detected until  the time $t$, then the
algorithm for the detection probability $A(\tau)$ at any later moment $\tau \geq
t$ should take into account the already achieved state $\phi_t$. However, {\em
what is} $\phi_t$?  By assuming that it is the output of the free evolution,
Kijowski extrapolates quite radically the classical picture. {\em The  particle 
moves along the straight line, knowing nothing about the  detector,  until  it 
makes a direct  hit.\/} The image is suggestive, but it neglects some important
facts. A characteristic property of quantum objects is that they diffract: so an
obstacle (detector) affects the propagation even if there was no absorption. If
we forget this, there is no quantum mechanics, so why to worry about the  ``time
 operator''? Yet, this is only the beginning of the troubles. 

If the  particle  propagates  in presence  of  a {\it waiting  detector\/} 
(screen, etc.), a new difficulty arises. Let's not forget  that  in   
quantum measurement theory, {\it no news is news.\/} If the detector shows  
a visible effect, the state of the microobject is {\it reduced.\/}  However,  if 
the detector {\it does not react, although it could,\/} the state of the
microobject is as well reduced (the reduction {\it by the absence of an
effect;\/}  see Dicke \cite{Dick}). The fact is essential for the well
known locality paradoxes (see Elitzur and Vaidman \cite{Eli}), for the
techniques of `seeing in the dark' \cite{Kwiat}, decisive also for 
the {\it time of event\/} of Blanchard and Jadczyk (BJ). They ask: {\em ``Is the
very presence of the counter reflected by the dynamics of the particles 
that pass the detector without being observed?''} (\cite{BJ96} p. 619, Sec.
2.3; see also \cite{segev}). The question, of course, is crucial for the waiting
screens. The positive  answer  to  BJ implies that $\phi_t$ varies not only due
to the free motion and/or diffraction on the screen but also due to the
progressive reduction process. The fact is considered as well in quantum
information and in cryptography (cf. an ingenious observation of the
experimental group in  Geneve, on the photon state which undergoes a non unitary
transformation, just because the photon {\it might have been absorbed}
\cite{Hutt}).

It thus seems, that the axioms about the {\em time of arrival\/} omit quite a
number of physical aspects. It brings little comfort that they give a unique
probability. On the contrary, it brings new difficulties. The screens used in
laboratories, quite obviously, must differ by some sensitivity parameters. So,
if the probability distribution in \cite{Ki74} is indeed unique, the problem
arises, where is the variety of the screen parameters?%
\footnote{An equivalent question was analyzed by Allcock with rather negative
results\cite{Allco}. It was also asked by Delgado and Muga in an opposite,
optimistic spirit\cite{De97}. Later on, a pertinen remark was made by Muga,
Baute, Damborenea and Egusquiza\cite{muga00}: Nevertheless, there is a clear
divorce between the daily routine of many laboratories where time-of-flight
experiments are performed, and the theoretical studies on the time of arrival,
which are based on the particle's wave function without recourse to extra
(apparatus) degrees of freedom. A number of ``toy models'' have been proposed
that include simple couplings between the particle's motion and other degrees of
freedom acting as clocks or stopwatches, but they do not include any
irreversibility\dots (c.f. p. 1, col. II, 1. 24--15 from the bottom), though
they no longer return to the problem in the published article\cite{Muga02}.}
Trying to answer that, one faces some more questions which might completely
frustrate the screen scenario behind the Kijowski operator (and behind the
related POV measures). 

Indeed, to have a unique probability distribution we must have an idealized
screen, for which the physical parameters become inessential. The options are
not many: ({\bf a}) If the screen permits the particles to tunnel unperceived
without any limitations, then the screen is ideal but nonexistent; ({\bf
b}) If the tunneling competes with the detection of the ``arrivals''
then the screen is not ideal and the probability distribution must depend on the
tunneling capacity. ({\bf c}) If the screen precludes completely the tunneling,
then the screen is indeed perfect. However, in both cases ({\bf b}) and ({\bf
c}) the particle evolution cannot be considered free before the detection
\cite{BM94,BJ96}. We thus face a mystery. What  is  the  physical  sense  of
Kijowski's probabilities and the related POV measures? 

In repeated discussions, several authors point out that the influence of a 
waiting screen (or other detectors) can be represented by a variant of an
optical model with a complex potential localized in the detector volume. For
the potential of an adequate structure\cite{complex1,complex2} the evolution of
the packet on one side of the screen (or detector boundary) is unaffected by the
existence of the detector, justifying the `free evolution hypothesis' in
\cite{Ki74,Ki99}; except that a part of the packet peacefully sinks into the
screen surface. According to Muga et al \cite{complex1}, this absorption can be
so perfect, that the influence of the complex potential can mimic every detail
of the free evolution, including the unavoidable back currents, though without
producing the reflected Fourier components\cite{complex1,complex2}. The gradual
reduction of the packet norm outside of the detector would then account for the
increasing absorption probability. While the mechanism deserves further study,
we doubt that it can confirm the Kijowski's probability distribution for the 
two-sided screens.  

The trouble is specially visible in the exceptional cases when the probabilities
of arrival should be obvious (at least, according to the common sense). We refer
to the case of {\it odd, normalized\/} packets $\phi(x,0)\equiv-\phi(-x,0)$
(cf. Muga and Leavens \cite{Leav00}, Sec. 8.8, p. 404). If the screen (detector)
placed at $x=0$ is represented by an absorbing complex potential $V(x)=V(-x)$
(with a narrow support, centered at $x=0$), the packet must remain odd for all
$t>0$, granting that $\phi_t$ vanishes forever at $x=0$ due to the destructive 
interference, and so does the probability current\cite{Leav00}. According to
the known statistical interpretation, it means that the absorption
probability should vanish. Looking for more details, Leavens had shown  
\cite{Leav98} that the Bohmian particle circulating in such packets never 
arrives at $x=0$. The same conclusion would hold in Nelson's stochastic quantum
mechanics \cite{Nelson} (the stochastic particle is repelled from the 
nodal points of $\phi_t$ \cite{Nelson,BoGo80}). However, the conclusion
does not indeed require the Bohm's or Nelson's theories. Through 
the entire history of quantum mechanics, the destructive interference was
decisive to determine the probabilities on the screens. This might change if the
screen (detector) precludes the free evolution (the possibility  neglected  in 
\cite{Ki74,Grot,De97,Muga98,De99,Leav00,Muga01,Muga02,muga00,complex1,complex2}
). Even so, the activated  screen  will  cut  the  communication between   the
packet components at $x<0$ and $x>0$. This is not the same as to separate
negative and positive momenta (which would be a nonlocal maneuver).

\begin{figure}
\includegraphics[width=6cm]{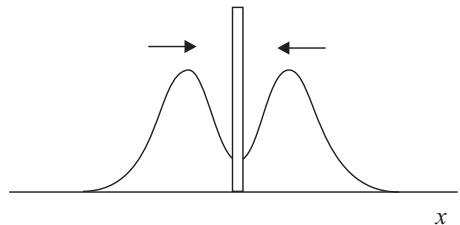}
\caption{A symmetric wave packet on both sides of the waiting screen. Either
the screen is transparent, or the shape of the packet must be affected by the
existence of the screeen even if there was no detection.}
\label{fig1}
\end{figure}

Note, that the trouble is not limited to the odd $\phi(x,t)$. 
To illustrate this, one can choose as well the symmetric, normalized packet
$\phi (x,0)=\phi (-x,0)$, which remains symmetric and normalizable for
$t>0$ (Fig.~\ref{fig1}). Once again, there is no need to worry about the
free evolution currents, since there are no such currents at $x=0$. If some
particles cross to the right and some other to the left there must be also
particles in the superposed states of crossing in both directions. Were the
evolution of the packet indeed free, then its both parts (for $x<0$ and $x>0$)
would conserve their norms (no absorption!) However, should the absorption occur
(e.g. due to the two sided analogue of the sinking mechanism described in
\cite{complex1}), the packet norm would be decreasing. A question arises: can 
this happen in such a way that the freely evolving packet is just multiplied by
an attenuating $c$-number factor $\lambda (t)$, $|\lambda (t)|<1$, without
losing its shape, so that the renormalized packet $\lambda (t)^{-1} \phi(x,t)$
would again evolve freely? This turns impossible, since then the packet
propagation would be affected in a non-local way, by adding to the free
Schr\"odinger's  Hamiltonian $H=p^2/2$ an imaginary time dependent constant 
(independent of $x$) instead of the localized screen potential $V(x)$. It means
that the vicinity of the screen must cause an essential change of
the packet shape, visible even after renormalizing $\phi_t$: so, the packet
evolution cannot be classified as ``free''\cite{complex2}. The same argument
shows that the proximity of a one sided, perfectly absorbing screen must as well
deform the packets.  

It thus looks that the complex models with imaginary peaks (or barriers) cannot
reproduce Kijowski's nor other distributions derived axiomatically from the free
evolution law. 

Strangely enough, a chance to rescue some operational aspects of
\cite{Ki74,Grot,De97,Ki99} might lie in an indifferent direction, i.e., close to
the option ({\bf a}). Perhaps, the acts of {\em crossing the screen\/} by
the microparticle should be understood as idealized events, developping in some
virtual (Platonic) reality, without the need of physical observation (cf. also
\cite{Leav00}, Sec. 8.8, p. 407). The possibility of detecting such unspoiled
events would occur in an asymptotic limit, not for strong, but inversely, for
very ``weak detectors'', for tiny screens, with almost negligible chances of
registering anything. In such situation, the {\em arrival\/} would not be a
synonym of the {\em detection.\/} The screen would seldom react, leaving
the free evolution practically unperturbed. Yet, should the (weak) detection
attempts be patiently repeated for ample sequences of initially identical wave
packets, the (conditional) probability distribution on the time which detection
moments are more and which are less probable. In this {\em scenario\/}, the
``crossing states''\cite{De97,Gian97,Muga98,De99,Unru99,Muga01}, would be a kind
of asymptotic states for an extremely weak screen-particle coupling (though
anchored at the finite time moments $t\in\mathbb{R}$). The results concerning
the time of arrival presented in\cite{Ki74,Grot} would be indeed the first order
approximation for the action of the detector in the {\em interaction
picture\/}-where the term {\em interaction\/} refers to the {\em weak\/}
particle-detector coupling. (Notice that this is merely a hypothesis, though
consistent with some elements which appear in Muga et al.; see
\cite{complex2}, p. 253, formula (A.8)). The influence of the stronger,
physically realistic detectors (or screens) on the evolution of the undetected
particles would correspond to the next approximation steps. The original
Kijowski formulae could hold approximately (in a weak coupling limit) if the
packet with positive momenta arrives to the screen 
from the far left, or else, a packet with negative momenta comes from the far
right; cf. \cite{De97} (the problem is, of course, how weak can be the
particle-screen coupling before we arrive at the quantum description of the
measuring devices, with all typical paradoxes)? In spite of all doubts, the idea
looks almost as a unique chance to give some operational sense to the 
known ``time of arrival'' distributions. However, it also reveals a gap in the
scriptures.  

If two packets, composed of the ``right'' and ``left movers'' arrive
simultaneously at the screen and are not coherent, the probabilities can be
simply added, justifying the well known POV measures\cite{Leav00}. If however 
they are two {\em coherent parts\/} of a single packet on both sides of the
detector, then it is not at all obvious how the weak coupling with the
screen can destroy the coherence (see again Leavens\cite{Leav98}, p. 844, col.
II, above (31)
\footnote{Commenting on Grot et al.\cite{Grot}, Leavens writes: {\em ``For the
theory based on Bohmian mechanics, interference between the two time-evolved
components of the pure state (19) has a dramatic effect on the distribution of
arrival times''(\dots) I find it a cause for concern that this important
ingredient of the theory apparently emerges as a consequence of the
regulation.\/} Indeed, it is difficult to disagree, even though the Bohm's
theory is lateral to the problem.}). It is even less obvious why a ``soft
screen'' should destroy the interference precisely between the {\em right} and
{\em left movers\/} (i.e., between the Fourier components). Both assumptions are
arbitrary, moreover, they are not identical. The tacit conviction about their
equivalence, exaggerates the role of the `Fourier thinking' in the problem
involving the space localization.

Quite evidently, the question about the ``time of arrival'' for a coherent, 
finite norm packet extending on both sides of the screen was never solved. As
it seems, the very concept of a {\em perfect screen\/} bears an intrinsic
{\em antinomy\/}. Either the screen is impenetrable (``strong''); then it
affects the packets even before the detection. Or the screeen (detector) is
extremely delicate (``soft''); then all operators and POV measures based on the
``interference ban'' are misleading. They were a useful intent, but the story
is not yet written.

\begin{center}
* * *
\end{center}

While the future of the subject is unknown, it becomes clear, that all intents
to obtain the {\em time observable\/} in the orthodox form of a self-adjoint
operator (in spite of the best stratagems to avoid the Pauli
theorem\cite{Pauli}) lead to a blind alley. The resulting operators are
typically plagued by some little but persistent difficulties which might look 
accidental; besides they all suffer some basic defect which seems common
for the whole family. As to the  Kijowski proposal of 1999 \cite{Ki99}, in the
first place, it is handicapped by an artificial form of $\Theta$, the continuous
spectrum operator, representing the {\it time\/} no better than
$\Xi(x)=\rm{sgn}(x)x^2/2$ imitates the harmonic oscillator! Observe, that
already the classical limit of $\Theta$: 
\begin{equation}
\Theta= -{z\over|p_z|}
\end{equation}
gives a false time value for 50\%  of  classical trajectories.

\begin{figure}
\includegraphics[width=4cm]{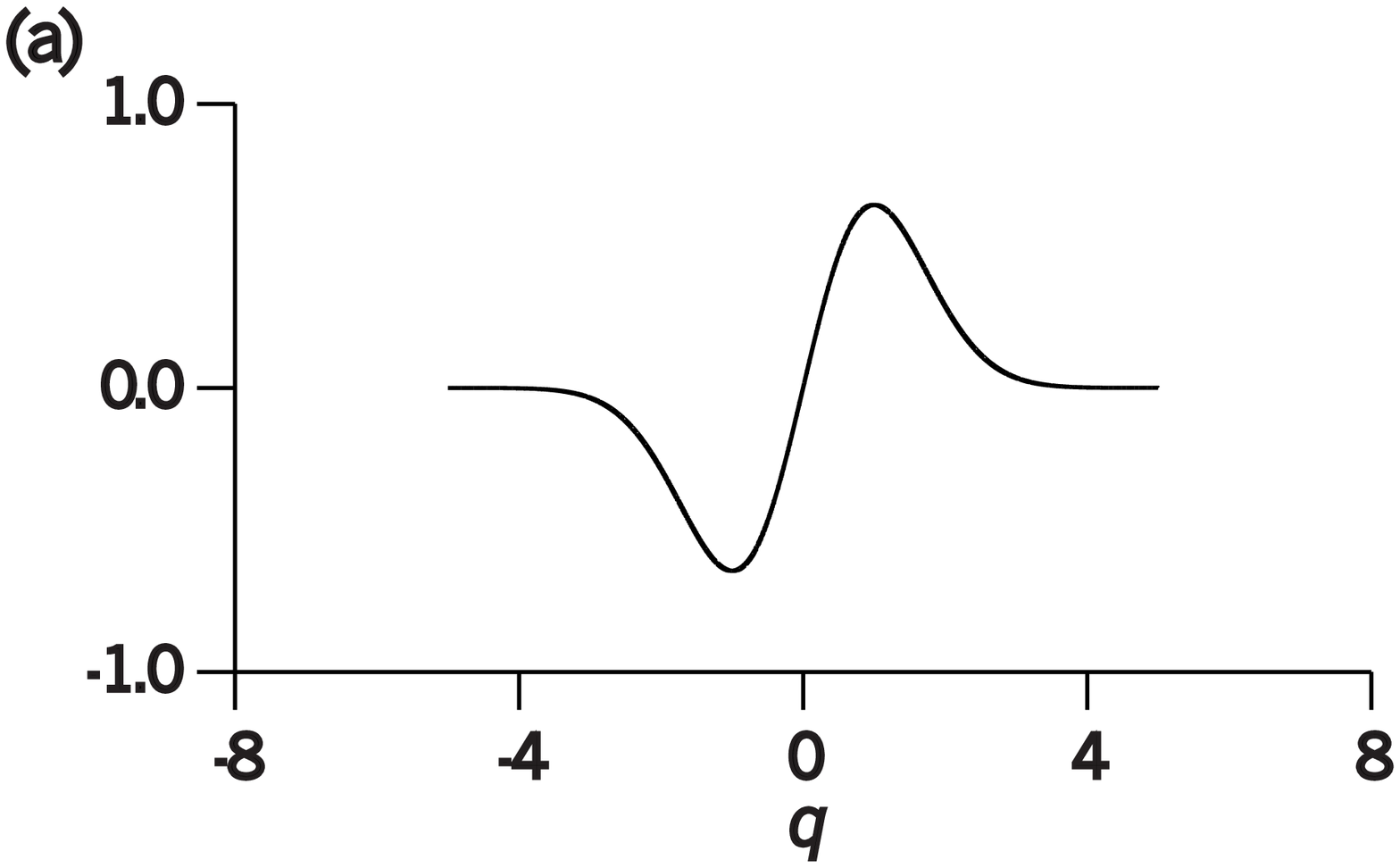}
\includegraphics[width=4cm]{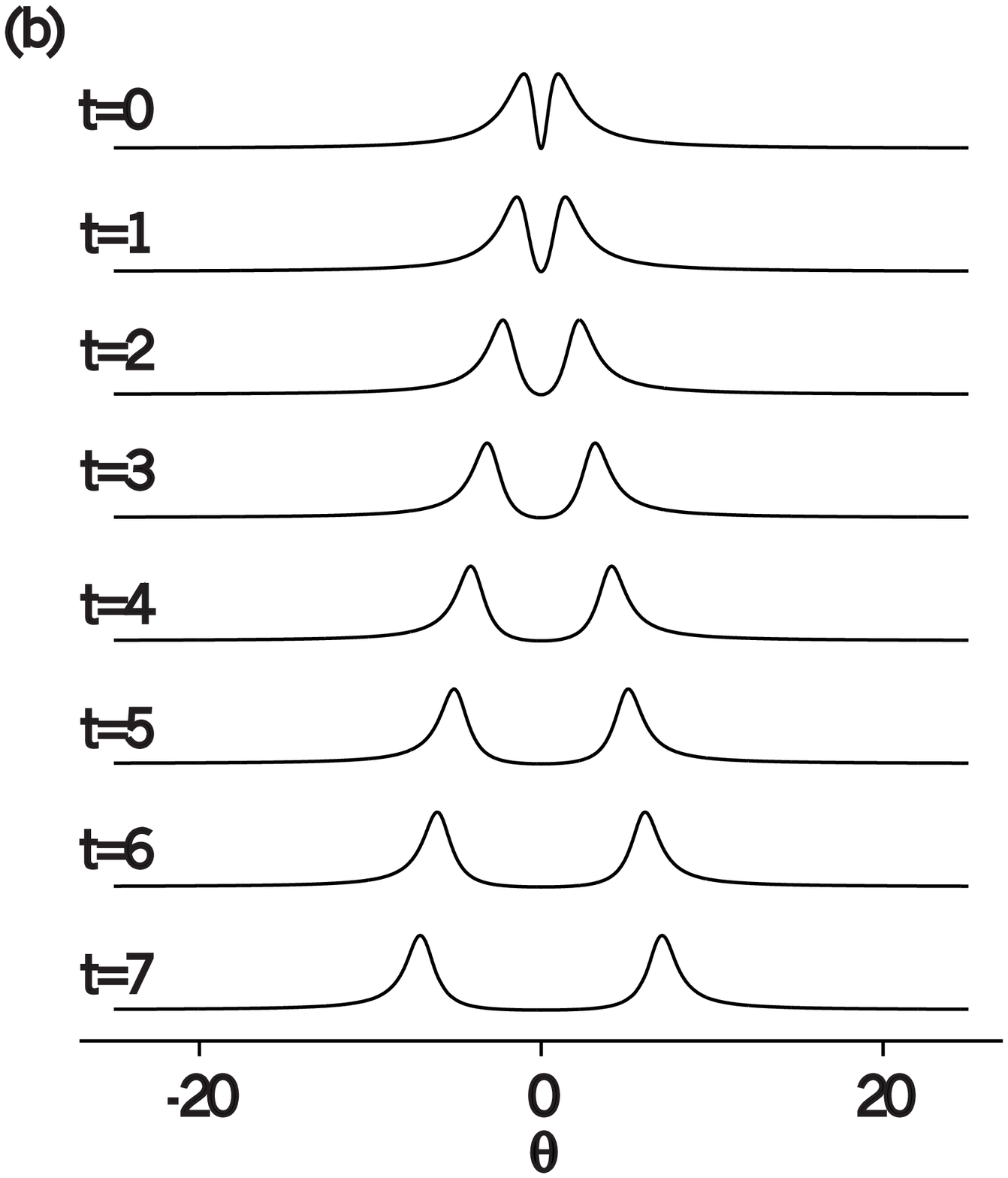}
\caption{The evolution of an odd packet seen on the $\theta$ axis. a) The 
initial packet in the $q$ representation. b) The Kijowski probability on the
$\theta$ axis forms two sharp peaks, ignoring the destructive interference. The
peaks subsequently separate, ignoring the covariance. The probability
distribution was corrected by POVM of Delgado and Muga, but the destructive
interference was not recovered.}
\label{fig2}
\end{figure}

To see what happens if one takes ``to the letter'' its quantum version (2)  
(see \cite{Ki99}), we have simulated the evolution of an odd wave packet,
vanishing all the time at $x=0$. The corresponding Kijowski probabilities 
do not seem to reflect the destructive interference at 0. Obeying rather the
superposition ban between the ``right'' and ``left'' movers, they produce 
two sharp peaks on the pseudo time axis (Fig.\ref{fig2}). As also turns obvious,
the probability formula (5) in \cite{Ki99} evolves covariantly with respect to
$\exp(-it\Xi)$ but not with respect to $\exp(-itH)$ (once again, $\Xi(p)$
represents $H$ no better than $\Xi(x)$ approximates the harmonic oscillator!). 

All this seems to support Allcock \cite{Allco} and Oppenheim et al.\cite{Unru99}
rather than the self-adjoint schools of Dirac \&  v. Neumann\cite{Dir,Neu}. 
We conclude that, in spite of its great inspiration, Kijowski's solution is too
biased by  the  orthodox doctrine. The attempts of Grot et al.\cite{Grot} do not
look more convincing. As it seems, the entire trend pays a price for excessive
idealizations: (1) for the lack of a physical scenario, (2) for overestimating
the role of the Fourier transforms in problems involving space and time
localizations, and (3) for introducing the probabilities which are valid only
when the experiment is never performed. 

Even if putting these questions aside, one faces an independent difficulty,
generic for all selfadjoint formalisms \cite{Dir,Neu}. The trouble consists  in 
the existence of ``absolute certainty states''. The fact, quite normal for the 
traditional (instantaneous) observables, here leads to unphysical 
conclusions. We shall discuss them taking {\em bona fide\/} the free evolution
background of Kijowski model.

Indeed, suppose, there exists a self-adjoint ``time operator'' 
$\hat{\rm{\bf t}}$  obeying the orthodox interpretations \cite{Dir,Neu}; so 
that the probability of the particle arrival in any time interval
$\Delta=[t_0,t_1]$ is given by the spectral projector $P(\Delta)$\cite{Ki99},
whose eigenstates imply the certainty. If  so, the orthogonal projectors
$P(\Delta)$ could be used to {\it assure the arrival\/} or to {\it grant the
absence.\/} We claim that this is a completely unrealistic conclusion. In fact,
let $\phi_0\in{\cal H}$, $P(\Delta)\phi_0=0$, be an initial state at $t=0$, 
granting that the microobject {\it will certainly not arrive\/} in an immediate 
future, i.e., in $\Delta=[0,t_1]$. Yet, if the  particle obeys the free
evolution equation for $t\in\Delta$ (as suggested in \cite{Ki74}) then,  
with rare exceptions \cite{Leav98,Heg98}, $\phi_t$ will immediately develop a
non vanishing current penetrating to the detector at $x=0$. We  find  it
entirely impossible to believe that this will traduce itself {\it
exclusively\/} into the tunneling effect, while the probability of the 
particle detection  will remain {\it exactly zero\/} in the entire interval
$\Delta$ (unless, of course, the detector is completely blind!). We henceforth
consider the absolute certainty states unphysical. The difficulty had been
foreseen by Aharonov et al.\cite{Aha98}, in form of an atypical time-energy
uncertainty relation $\Delta t>1/E$ (where $E$ is the initial kinetic energy),
implying that the ``certainty'' offered by the spectral projectors is
illusionary.

\begin{figure}
\includegraphics[width=3.8cm]{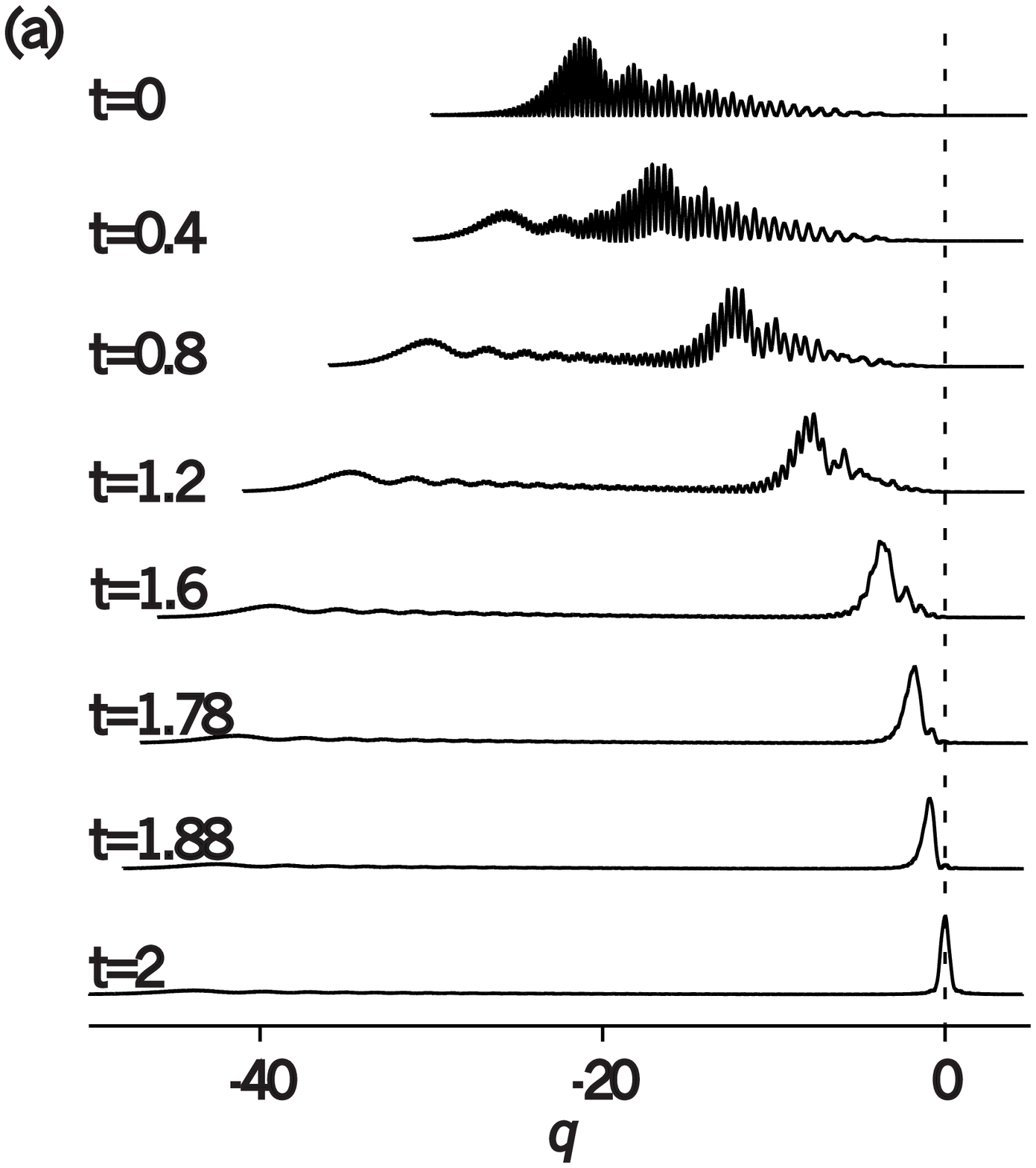}
\hspace{1pc}
\includegraphics[width=4cm]{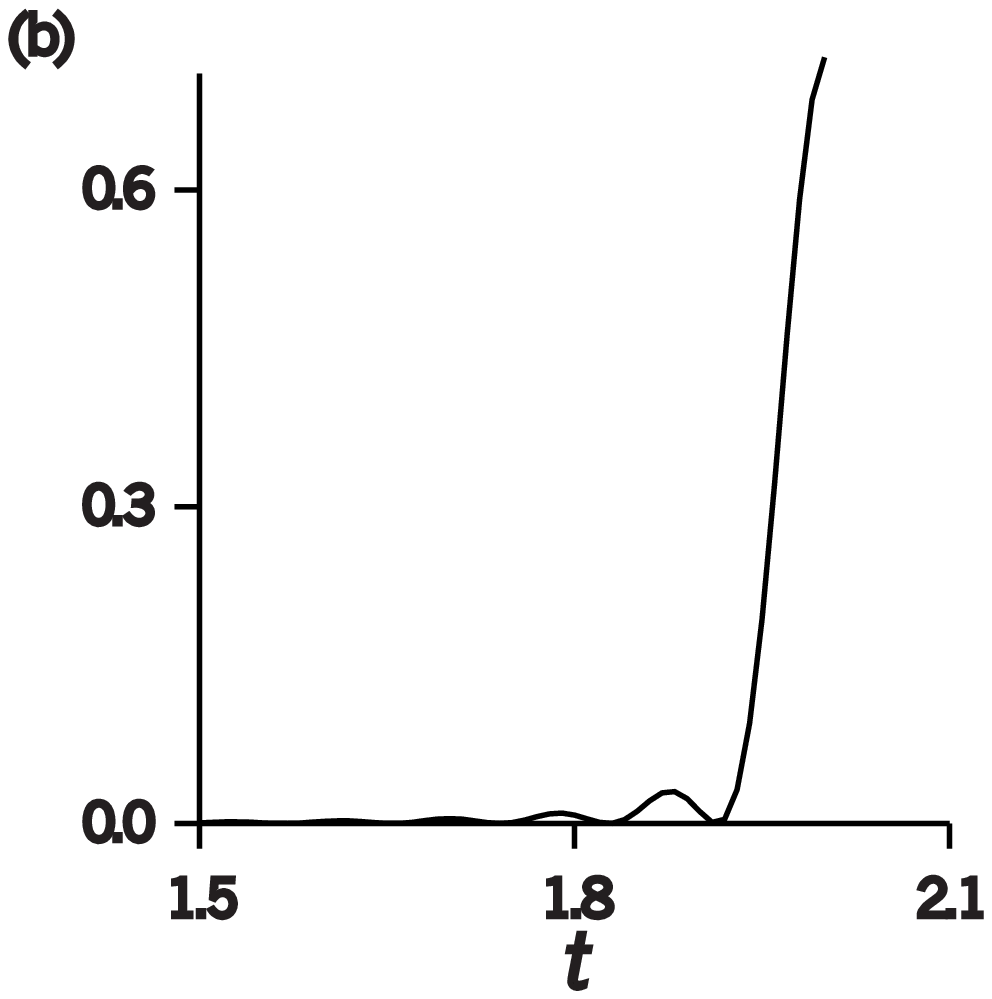}
\caption{(a) The position representations for the wave packet initially forming 
a sharp step on the $\theta$ axis in the narrow interval [2,200001]. For $t\to2$
more than $1/2$ of the packet is far from 0. Moreover the packet develops a
visible tail at $x=0$. b) The tail absolute values at $x=0.003$ (behind the
detector) for $t<2$, $t\to2$. We refuse to believe that the phenomenom will
traduce itself only into the packet tunneling without absorption.}
\label{fig3}
\end{figure}

To clarify these doubts, we have carried out some numerical tests. For
simplicity, we consider  the  particle  in  1-space  dimension.  Our detector is
a ``sensitive  eye''  placed  at  $x=0$. The detector can be switched on and off
at  will  (eye  open/closed)  in  any desired time interval. Since the correct
theory must resist unfriendly tests, we have chosen our initial wave function
$\phi_0$ to be just a step  function in $\theta$-representation, vanishing 
outside  of  a  narrow  interval  $[\theta_1=2, \theta_2=2.00001]$. According to
\cite{Ki99} it should give an absolute  certainty that the particle {\it will 
not  be  registered\/}  for  $t<2$, and that it  {\it must arrive\/} in 
$2<t<2.00001$.  The  sequence  of  graphics  on  Fig.~\ref{fig3} represents the
free evolution of our wave function in the position representation as $t$
approaches $t_1=2$. As  can  be  seen, around half of the packet is far away
from the detector for $t$  close  to 2, contradicting the ``absolute 
certainty'' of the particle arrival between $t_1=2$ and  $t_2=2.00001$ (cf.
also \cite{Unru99}). The other problem is even deeper. While approaching  the 
time moment  $t_1=2$ from below, $\phi_t$ develops an inreasing tail around
$x=0$. Should  the  window  of awareness be e.g. [1.99,2.00], it  is  hard  to 
believe  in  the  absolute impossibility of the particle detection in this
window, i.e. {\it before\/} the time allowed by Kijowski projector. 

Note, that the situation would not change for any different self adjoint,
positively defined Hamiltonians\cite{Leo00,Bau00}, since our graphics illustrate
simply the general {\em no go theorem\/} of
Hegerfeldt\cite{Heg98,Heger,Skulim1,Skulim2,Skulim3}.

In spite of all objections, an intresting expression of these facts are 
the ``states of arrival'' $|\pm,t\rangle$ (also {\em crossing states}), studied
in\cite{De97,Gian97,Muga98,De99,Unru99,Leav00,Muga02,Skulim1,Skulim2,Skulim3,%
leo97,gianni98,gala}, forming an overcomplete, nonorthogonal basis. Note that
the mappings $t\rightarrow|\pm,t\rangle$ establish indeed a {\em fuzzy
structure\/} on the real time axis. In the recent research, such structures are
often defined by mapping the points of classical differential manifolds into the
families of non-orthogonal {\em coherent states}\cite{Oqsp,Ocoh} which form the
overlapping, {\em fuzzy images\/} of the originally distinguishable points (for
a different approach, see \cite{dragan01,Bala02,Denjoe,Pavel}). The sense of the
``crossing states'' seems analogous. Each falling particle watches the time
axis; it must chose a moment $t$ to hit the screen surface. However, instead of
the continuum of sharply defined time points, the particle `{\em sees}' a family
of fuzzy events in form of the {\em crossing states\/} $|+,t\rangle$ and/or
$|-,t\rangle$ with nontrivial overlaps. As a result of this {\em fuzzy
vision,\/} the particles
cannot be instructed to hit the screen in sharply defined time intervals: i.e.,
the ensemble preparation does not admit the {\em absolute certainty states.\/}
The same phenomenon seems generic for the time localizations of other
quantum events.

Quite obviously, the  waiting detectors form a new class of measurements, which
have rather little to do with the traditional Dirac-v.~Neumann observables.
Though all this concerns a non-relativistic theory, it might mean a warning for
the `Euclidean' space-time quantizations\cite{reis,marolf} where the space and
time localizations receive an equal status (see also the discussions by
Le\'on\cite{leo97} and Giannitrapani\cite{gianni98}).

Note, that this negative conclusions might also imply some good news. Indeed,
imagine a simple experiment which consists in registering the time moment in
which an unstable particle decays. The result is a real number (the {\it decay
time}); yet, an attempt to describe it  in  frames of the orthodox scheme (i.e.,
as an eigenvalue of a self-adjoint {\it time operator\/}) would lead to a wrong
conclusion about the existence of initial states for which the moment of the
decay can be predicted with certainty! Should this be true, the consequences
could be quite dramatic. They would include a suspense story about a suitcase
full of radioactive atoms smuggled safely through the custom control, with all
atoms programmed to decay tomorrow! So, perhaps, we should not regret that the
{\it time of events\/} does not obey the orthodox axioms of Dirac and
v.~Neumann? 

The stimulating discussions of M. Przanowski and M. Skulimowski are appreciated.
The support from Conacyt grant 32086E is acknowledged.

\end{document}